# Formation of buried conductive micro-channels in single crystal diamond with MeV C and He implantation

F. Picollo[1], P. Olivero[1]*, F. Bellotti[2], J. Pastuović[3], N. Skukan[3],

A. Lo Giudice[1], G. Amato[2], M. Jakšić[3], E. Vittone[1]

[1] Experimental Physics Department, INFN Sezione di Torino, Centre of Excellence "Nanostructured Interfaces and Surfaces" (NIS), Università di Torino, via P. Giuria 7, 10125 Torino, Italy

[2] Quantum Research Laboratory, Istituto Nazionale di Ricerca Metrologica, Strada delle Cacce 91, 10135 Torino, Italy

[3] Laboratory for Ion Beam Interactions, Ruđer Bošković Institute, Bijenička 54, HR-10000 Zagreb, Croatia

* corresponding author (olivero@to.infn.it)

## Abstract

As demonstrated in previous works, implantation with a MeV ion microbeam through masks with graded thickness allows the formation of conductive micro-channels in diamond which are embedded in the insulating matrix at controllable depths [P. Olivero et al., Diamond Relat. Mater. 18 (5-8), 870-876 (2009)]. In the present work we report about the systematic electrical characterization of such micro-channels as a function of several implantation conditions, namely: ion species and energy, implantation fluence. The current-voltage (IV) characteristics of the buried channels were measured at room temperature with a two point probe station. Significant parameters such as the sheet resistance and the characteristic exponent ($\alpha$) of the IV power-law trend were expressed as a function of damage density, with satisfactory compatibility between the results obtained in different implantation conditions.

**Keywords**

Diamond crystal, ion implantation, electrical conductivity

## Introduction

Since the pioneering works by Vavilov and Hauser in the 70s [1, 2], the modification of the electrical transport properties of diamond when subjected to ion damage has been widely investigated. After the early demonstrations that the electrical properties of ion-implanted diamond layers were similar to those of amorphous carbon [3], the hopping conduction mechanisms in C-implanted samples were studied by Prins [4, 5]. The influence of target temperature during Ar and C implantation was investigated by Sato et al. [6, 7], while the first studies on polycrystalline samples grown by chemical vapor deposition were conducted by S. Prawer et al. [8]. The effect of ion fluence was systematically studied in samples implanted with C/Xe and B in a series of works by S. Prawer et al. [9, 10] and F. Fontaine et al. [11], respectively. An extensive current-voltage (IV) characterization in temperature of Xe-implanted diamond was carried by A. Reznik et al. [12, 13], allowing the extraction of a number of characteristic energies for hopping conduction sites [14].

A widely accepted interpretation of the modification of the electrical properties of damaged diamond is based on the existence of two damage regimes. When the damage density is below a critical threshold (often referred as "graphitization threshold" or "amorphization threshold"), the conduction mechanism can be described with the variable range hopping within a sparse network of defective $sp^3$ and $sp^2$ sites, while at higher damage densities a continuous network of $sp^2$-bonded defects is formed, which leads to the permanent graphitization of the structure upon thermal annealing and to the subsequent appearance of metallic conductivity.

It is worth stressing that all of the above-mentioned works were performed by damaging the samples in superficial regions with heavy ions at energies in the $10^1$-$10^2$ keV range. Remarkably, the possibility of fabricating sub-superficial channels with damage-related conductivity induced by MeV ions has been explored in a limited number of works, both in the sub-graphitization [15] and in the graphitization [16] regimes. Recently, we demonstrated that it is possible to created buried conducting micro-channels in diamond with MeV ion implantation through variable-thickness masks, allowing the emergence of the channels at their endpoints, thus in electrical contact with the sample surface electrodes [17]. Moreover, in a recent work such structures were characterized with IV measurements in temperature, allowing the elucidation of variable-range-hopping conduction mechanisms in as-implanted samples [18].

In the present work, we report about the systematic IV characterization at room temperature of buried micro-channels as a function of several implantation conditions: ion species and energy (6 MeV C, 1.8 MeV He), and implantation fluence ($3 \cdot 10^{15}$ - $4 \cdot 10^{16}$ cm$^{-2}$ and $3 \cdot 10^{16}$ – $6 \cdot 10^{17}$ cm$^{-2}$, respectively).

## Experimental

The process of damage induced by MeV ions in matter occurs mainly at the end of ion range, where the cross section for nuclear collisions is strongly enhanced, after the ion energy is progressively reduced by electronic interactions occurring in the initial stages of the ion path [19]. In Fig. 1 we report the damage density profiles of 6 MeV C and 1.8 MeV He ions as resulting from the Monte Carlo simulation with SRIM code [20]. The curves were calculated by setting a value of 50 eV for the atom displacement energy in the diamond lattice [21, 22]. As visible in Fig. 1, the high damage density at the end of range of the ions determines the formation of narrow damaged layers at depths of ~2.7 μm and ~3.2 μm below the sample surface for 6 MeV C and 1.8 MeV He, respectively. The damage density induced by carbon ions is significantly higher with respect to helium ions, with an integrated number of vacancies per ion of 299 compared to 49, respectively.

In order to electrically connect the endpoints of the channels to the surface contacts, a three-dimensional masking technique was employed to modulate the penetration depth of the ions from their range in the unmasked material up to the sample surface with increasing thickness of stopping material. The basic concept is shown schematically in the inset of Fig. 2, while further details can be found in [17].

The samples consist of synthetic single crystal diamonds classified as type Ib, with a substitutional nitrogen concentration comprised between 10 and 100 ppm. The diamonds were produced with high pressure high temperature (HPHT) synthesis by Sumitomo, and their size is 3×3×1.5 mm. The samples are cut along the 100 crystal direction and they are optically polished on the two opposite large faces. After the masking process, the samples were implanted at room temperature with two different ion species, namely carbon and helium. The implantation of 1.8 MeV He ions was performed at the micro-beam line on the AN2000 accelerator facility of the INFN Legnaro National Laboratories (INFN-LNL). 6 MeV C ions were implanted at the microbeam line of the Laboratory for Ion Beam Interaction of the Ruđer Bošković Institute (LIBI-RBI). In both the implantations the ion beam was focused to a micrometric spot, and the ion current was ~0.8 nA and ~0.3–5 nA for the C ion and He ion respectively. The emission of backscattered ions or characteristic x-rays from the metal electrodes was employed to monitor the implantation fluence in real time respectively during C and He implantation, after a suitable calibration with a Faraday Cup. The accuracy of the fluence evaluation is estimated of the order of 20%, whereas the uncertainty is negligible.

Before ion implantation, the sample was masked with semi-spherical contacts by means of a standard gold wire ball bonder commonly used for the contacting of microchip devices. In order to improve the adhesion of the gold contacts, Cr-Au adhesion areas were evaporated on the sample surface through a mask, as shown schematically in the inset of Fig. 2. The Cr and Au adhesion layers are <100 nm in thickness, therefore they do not affect significantly the depth at which the damaged layer is formed in diamond; on the other hand, the maximum thickness of the gold contacts is of the order of 50 μm in their central regions, which is more than enough to fully stop the incident ions. Fig. 2 shows an optical image in transmission of a sample implanted with 1.8 MeV He and 6 MeV C at increasing fluences, in the ranges ($3 \cdot 10^{16}$ – $6 \cdot 10^{17}$ cm$^{-2}$) and ($3 \cdot 10^{15}$ – $4{\cdot}10^{16}$ cm$^{-2}$), respectively. The optically opaque implanted channels are clearly visible, together with the Cr/Au adhesion layers and the semi-spherical contacts.

The electrical characterization was performed with a Semiconductor Parameter Analyzer (4145B Hewlett Packard) connected to a probe station (Alessi REL-4500). The IV curves were measured in "voltage source" mode by positioning the two probe tips on the channels end-points and making a sweep in the (-100 – 100 V) range. Test IV measurements were performed between contacts that weren't connected through micro-channels, allowing an effective check that the measured conductivity is only relevant to the sub-superficial layers and not to spurious surface conductivity.

**Results and discussion**

Fig. 3 shows the measured IV curves; as expected, the IV characteristics are symmetrical with respect to the applied voltage polarity, therefore only the data in the 0 – 100 V range are reported. Firstly, it is worth noticing that the IV curves systematically exhibit a higher conductivity for channels implanted at higher fluences with both He and C ions. The test measurement between unconnected contacts (not reported here) yield a current signal comparable with the instrumental noise (~$10^{-10}$ A), therefore we can conclude that the conduction mechanism is entirely to be attributed to buried channels, with negligible contributions from surface conductivity. Secondly, a linear IV characteristic in the whole 0 – 100 V range is observed only for channels implanted at low fluence, both for He and C implantations (see Fig. 3a), while a super-linear trend is systematically observed from channels implanted at higher fluences (see Fig. 3b).

In order to determine the transport mechanism at low electric fields as function of different irradiation conditions, a suitable scaling procedure of the electrical resistance of the channels and of the implantation fluences has been adopted.

Firstly, since implantations performed with different ion species and energies determine significantly different damage profiles in the material (as shown schematically in Fig. 1), the fluences were re-scaled by multiplying their values with factors $I_i$, defined as:

$$I_i \equiv \int_0^d p_i(z)dz \tag{1}$$

where subscript index i refers to the ion energy end species, d is the penetration range of the ions in diamond, and $p_i(z)$ is the damage linear density, expressed as the number of created vacancies per unit of penetration depth per incoming ion, i.e. the quantity plotted in Fig. 1. Therefore, since $I_i$ can be regarded as the total number of vacancies generated by a single ion (i.e. 299 vacancies for 6 MeV C ion and 49 vacancies for 1.8 MeV He ion), the product $\nu_{m,i}* \equiv \Phi_{m,i} \cdot I_i$ (where $\Phi_{m,i}$ is the fluence of the $m^{th}$ implantation for the $i^{th}$ ion species and energy) can be interpreted as the total number of vacancies generated in the implantation per surface unit. Such an interpretation is valid only under the assumption that the local density of vacancies at a given depth (*z*) in the material, i.e. $\nu_{m,i}(z)$, is directly proportional to the number of implanted ions, hence:

$$\nu_{m.i}(z) = \Phi_m \cdot p_i(z) \quad \Rightarrow \quad \nu_{mi}* = \int_0^d \nu_{mi}(z)dz = \Phi_m \cdot I_i \tag{2}$$

The above-mentioned assumption is indeed rather simplicistic, since it does not take into account non-linear processes occurring in the damage process, such as self-annealing, ballistic annealing and defect interaction. Nonetheless, the variable $\nu_i$* can be regarded as a first-order parameter to re-scale fluences relevant to implantation processes performed with different ion species and energies.

Concerning the electrical observable, the resistance of the channels in the low-field regime was evaluated from a linear fit of the IV data in the 0 – 1 V range, where all curves exhibit a linear trend. Moreover, a similar scaling process can be applied to the value of the channel resistivity, which is defined as $\rho = \frac{R \cdot w \cdot t}{L}$, where R is the channel resistance, and w, t and L are its width, thickness and length, respectively. Again, since the buried channel is defined in the depth direction by a strongly non-uniform damage profile (see Fig. 1), the sheet resistance can be defined as $\rho^* \equiv \frac{\rho}{w} = \frac{R \cdot t}{w}$. Remarkably, while the $\rho[\Omega \cdot \mathrm{cm}]$ vs $\Phi[\mathrm{ions} \cdot \mathrm{cm}^{-2}]$ plots yield to inconsistent results when data relevant to different implantation processes are merged, the relevant $\rho^*[\Omega/\square]$ vs $\nu^*[\mathrm{vacancies} \cdot \mathrm{cm}^{-2}]$ plots result in satisfactory consistency, as shown in Fig. 4. In the above-mentioned plot, the sheet resistance exhibits a monotonous decrease in the $10^6 - 10^9$ $\Omega/\square$ range at increasing values of damage surface density in the $10^{18}$ – $10^{20}$ vacancies $\mathrm{cm}^{-2}$, with satisfactory compatibility between data relevant to different implantation processes.

Such scaling process can be employed also for the comparison of the above mentioned non-linearity of the IV characteristics as a function of implantation fluence. The field-enhanced transport mechanism in channels implanted at high fluences can be modelled with the power-law dependence $I \propto V^{\alpha}$. In Fig. 3 the fitting functions are reported with continuous lines, showing satisfactory agreement with the experimental data, both in the linear and super-linear regimes. Similarly to what reported above, the $\alpha$ vs $\Phi[\mathrm{ions} \cdot \mathrm{cm}^{-2}]$ plots yield to inconsistent results between the two implantation processes, while the scaled α vs ν*[vacancies·cm$^{-2}$] plot consistently shows a progressive increase of the *α* parameter as a function of increasing damage density, from ~1 (corresponding to linear behavior) up to ~1.45. Results of previous [18] conductance measurements in the temperature range 250-690 K carried out on a single channel are compatible to the systematic data reported here, as shown in Fig.5. Moreover, they suggest that such super-linear (α>1) behavior at moderate electric field (> 500 V cm$^{-1}$) is to be ascribed to a space charge limited transport mechanism influenced by the trap type and distribution in the bandgap [23]. The behavior of α as function of the damage density as shown in Fig. 3 seems to suggest the existence of a critical damage threshold occurring between $5 \cdot 10^{18}$ and $1 \cdot 10^{19}$ vacancies cm$^{-2}$, and relevant to a transition from ohmic to space charge limited current transport mechanism in the moderate electric field region.

**Conclusions**

We reported about the fabrication and electrical characterization of buried conductive micro-channels in single crystal diamond with MeV ion induced damage. The analysis of the IV characteristics yielded significant parameters describing the conduction properties of the channels, namely the sheet resistance and the characteristic exponent (α) of the power-law IV trend, which were expressed as a function of damage density. Satisfactory compatibility between the results obtained in different implantation conditions (ion species and energy) was achieved by suitably scaling the implantation fluences with parameters relevant to the different damage profiles. A peculiar evolution of the α exponent as a function of the implantation fluence was identified, which will be investigated in further details in future works.

**Acknowledgments**

The work of P. Olivero is supported by the “Accademia Nazionale dei Lincei – Compagnia di San Paolo” Nanotechnology grant, which is gratefully acknowledged.

**Figure captions**

Fig. 1: damage density profiles of 6 MeV C (dashed red line) and 1.8 MeV He (continuous blue line), expressed in linear density of generated vacancies per incoming ion, as resulting from the Monte Carlo simulation with SRIM code [20]; the high damage density at the end of ion range is clearly visible.

Fig. 2: optical microscopy image in transmission of the implanted sample; the opaque micro-channels (labeled as “1-7” and “A-F” for 1.8 MeV He and 6 MeV C implantations, respectively) are clearly visible, together with the adhesion layers and the semi-spherical gold contacts. In the inset a cross-sectional schematic of the three-dimensional masking process is drawn; more details can be found in [17].

Fig. 3: IV curves of channels implanted with 1.8 MeV He and 6 MeV C ions at (a) low and (b) high fluences; while in the former case the IV curves are linear, in the latter case the curves follow a super-linear power-law dependence; experimental data are reported in dots, while fitting curves are reported in continuous lines.

Fig. 4: log-log plot of the sheet resistance of the channels as a function of the damage surface density, for 1.8 MeV He (red dots) and 6 MeV C ions (black squares); the monotonous decrease exhibits a pseudo-exponential behavior.

Fig. 5: characteristic exponent $\alpha$ of the power-law trend of the IV curve $I \propto V^{\alpha}$, as a function of the damage surface density, for 1.8 MeV He (red dots) and 6 MeV C ions (black squares); the monotonous increase exhibits a significant transition in the $5{\cdot}10^{18}$ – $1{\cdot}10^{19}$ vacancies $cm^{-2}$ range; the data are compatible with what reported in Ref. [18] for the IV characterization in temperature of a single channel (green dot).

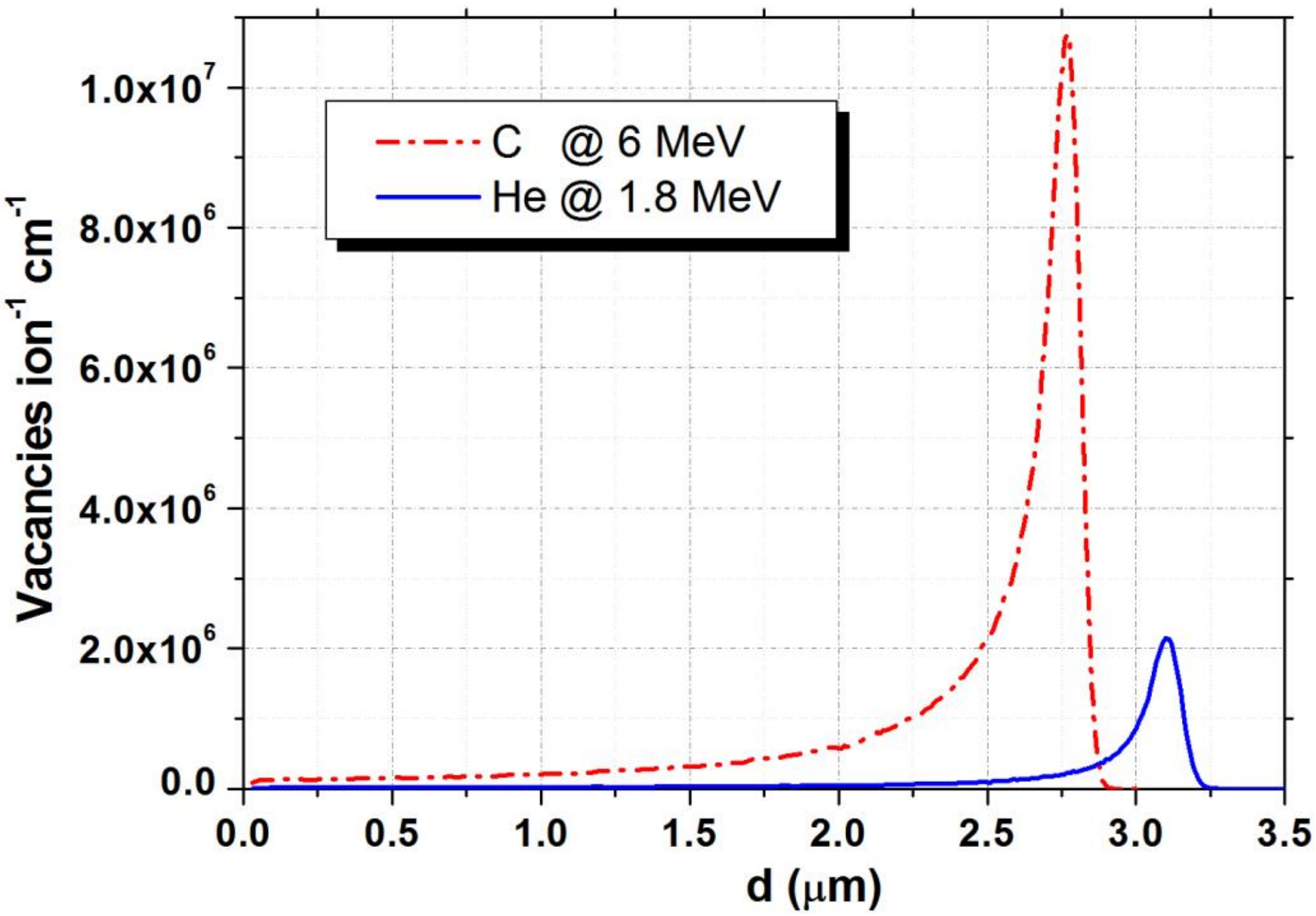

C @ 6 MeV
He @ 1.8 MeV
Vacancies ion$^{-1}$ cm$^{-1}$
1.0x10$^{7}$
8.0x10$^{6}$
6.0x10$^{6}$
4.0x10$^{6}$
2.0x10$^{6}$
0.0
0.0 0.5 1.0 1.5 2.0 2.5 3.0 3.5
d (μm)

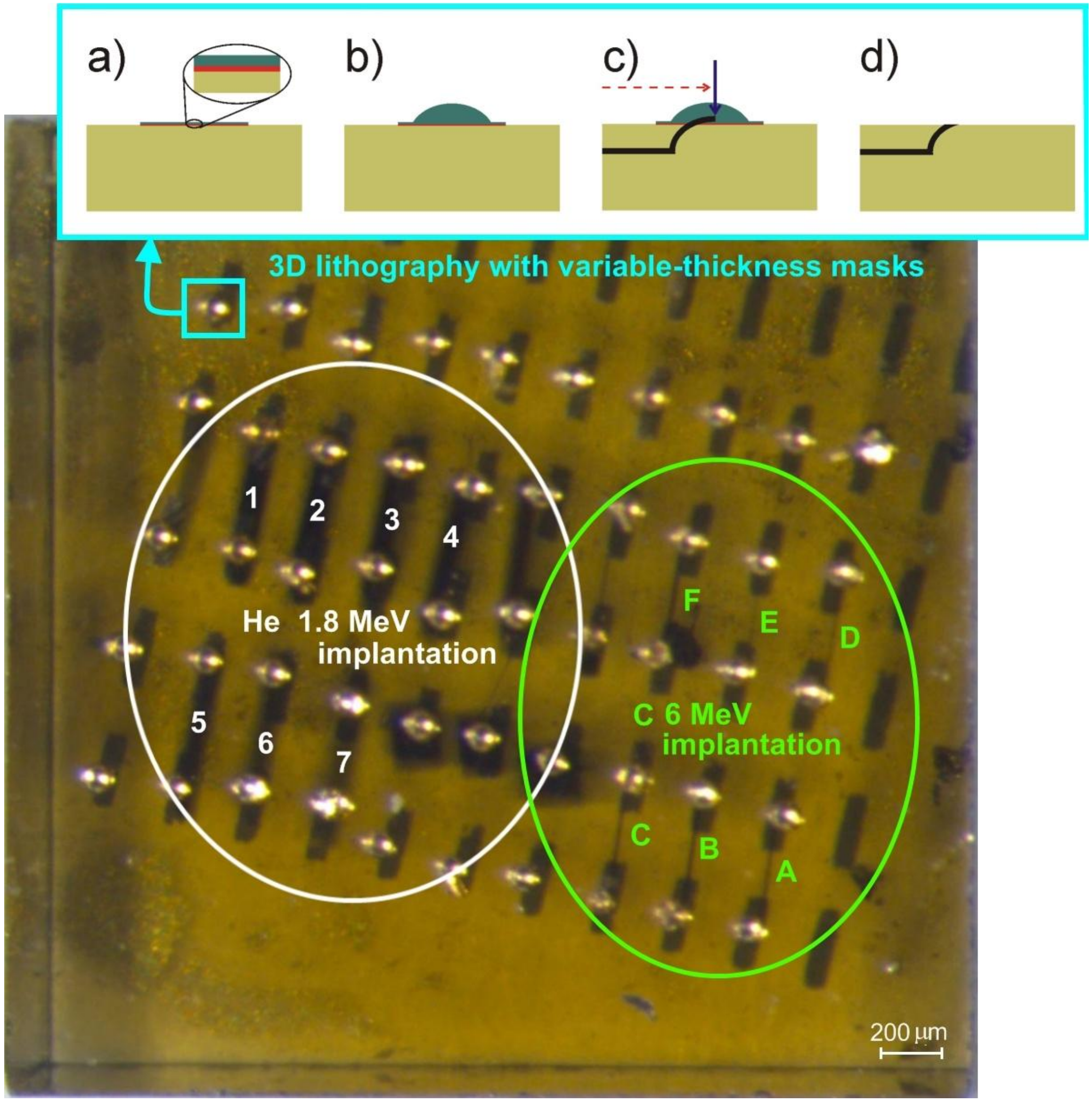

a)
b)
c)
d)
3D lithography with variable-thickness masks
1
2
3
4
He 1.8 MeV
implantation
5
6
7
F
E
D
C 6 MeV
implantation
C
B
A
200 µm

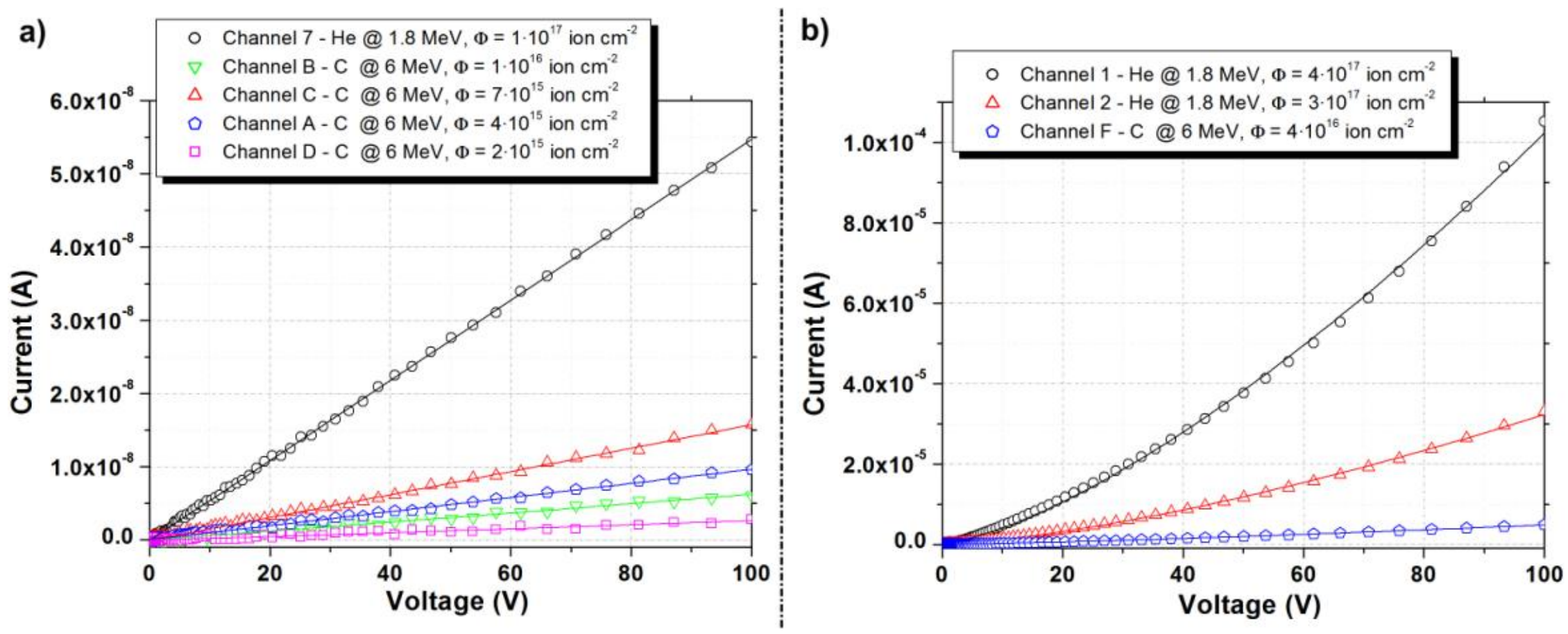

a)
Channel 7 - He @ 1.8 MeV, Φ = 1·10^17 ion cm^-2
Channel B - C @ 6 MeV, Φ = 1·10^16 ion cm^-2
Channel C - C @ 6 MeV, Φ = 7·10^15 ion cm^-2
Channel A - C @ 6 MeV, Φ = 4·10^15 ion cm^-2
Channel D - C @ 6 MeV, Φ = 2·10^15 ion cm^-2
Current (A)
Voltage (V)
b)
Channel 1 - He @ 1.8 MeV, Φ = 4·10^17 ion cm^-2
Channel 2 - He @ 1.8 MeV, Φ = 3·10^17 ion cm^-2
Channel F - C @ 6 MeV, Φ = 4·10^16 ion cm^-2
Current (A)
Voltage (V)

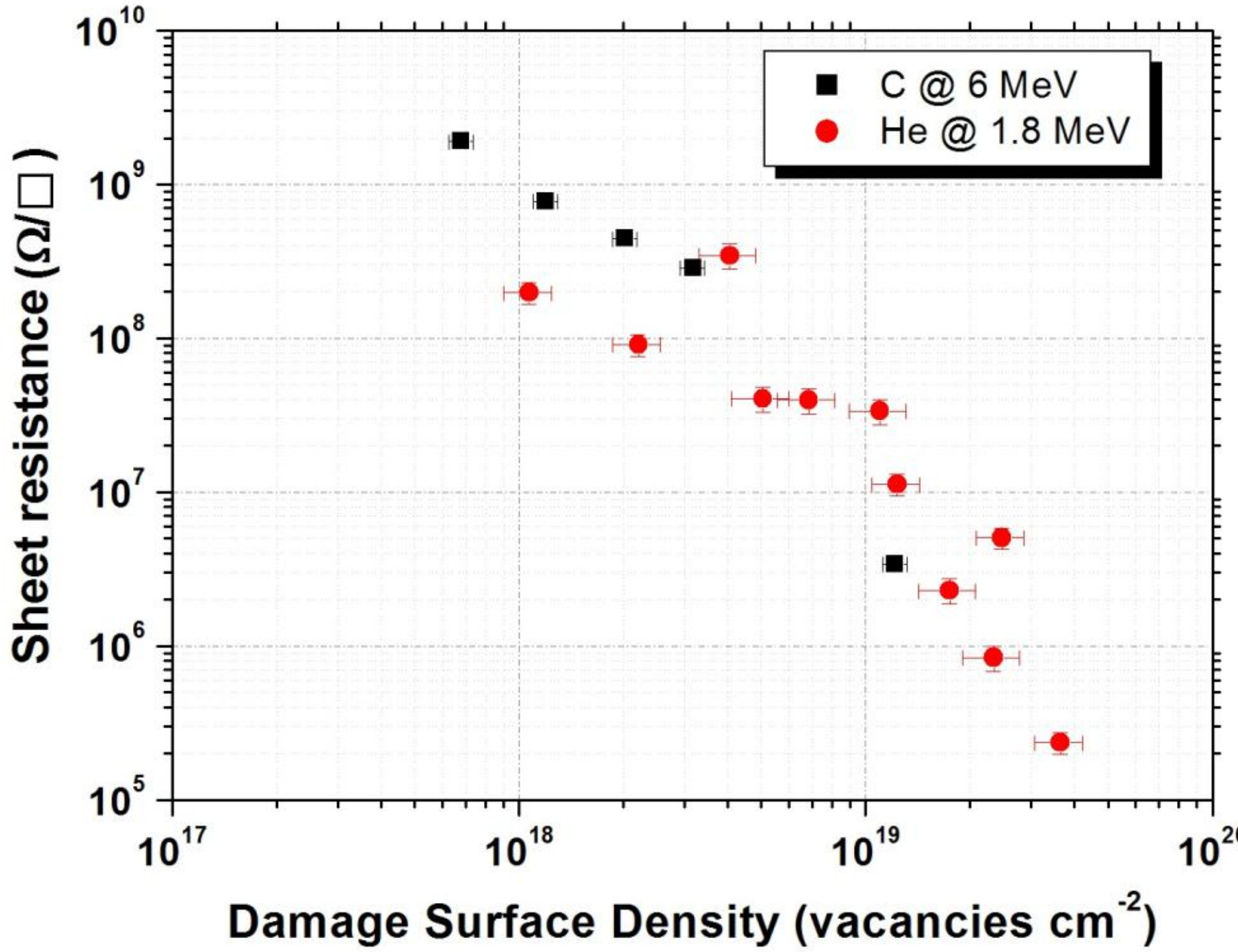

C @ 6 MeV
He @ 1.8 MeV
Sheet resistance (Ω/□)
Damage Surface Density (vacancies cm$^{-2}$)
$10^{10}$
$10^{9}$
$10^{8}$
$10^{7}$
$10^{6}$
$10^{5}$
$10^{17}$
$10^{18}$
$10^{19}$
$10^{20}$

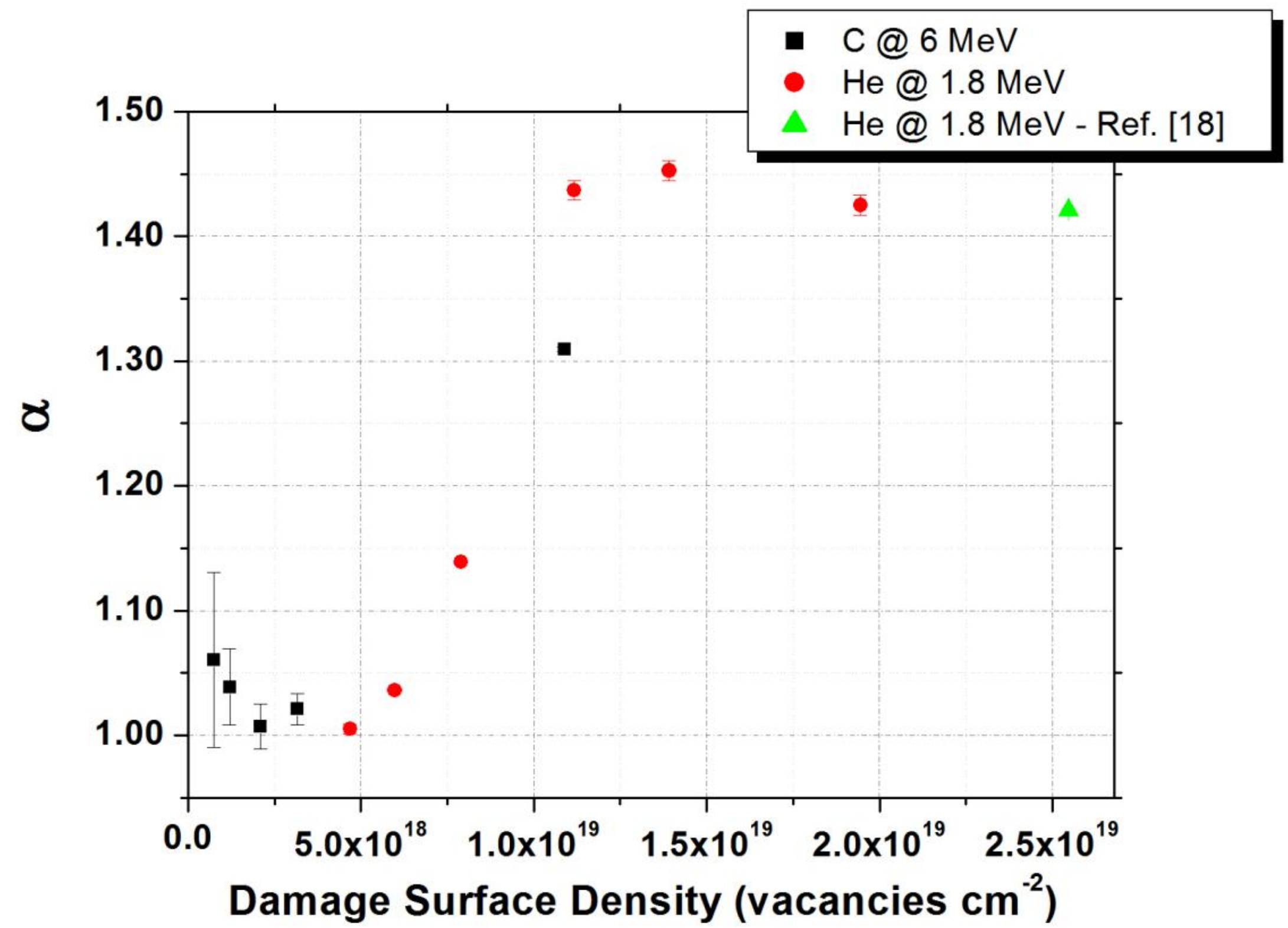

C @ 6 MeV
He @ 1.8 MeV
He @ 1.8 MeV - Ref. [18]
1.50
1.40
1.30
1.20
1.10
1.00
α
0.0
5.0x10^18
1.0x10^19
1.5x10^19
2.0x10^19
2.5x10^19
Damage Surface Density (vacancies cm^-2)